\documentclass[prd, twocolumn, superscriptaddress]{revtex4-2}
\usepackage{amsmath}
\usepackage{amssymb}
\usepackage{xcolor}
\usepackage{graphicx}
\usepackage{soul} % strike over chars with \st
\begin{document}

\title{Limits to the observation of the Unruh radiation via first-quantized hydrogen-like atoms}

\author{Riccardo Falcone}
\affiliation{Department of Physics, University of Sapienza, Piazzale Aldo Moro 5, 00185 Rome, Italy}

\author{Claudio Conti}
\affiliation{Department of Physics, University of Sapienza, Piazzale Aldo Moro 5, 00185 Rome, Italy}
\affiliation{Research Center Enrico Fermi, Via Panisperna 89a, 00184 Rome, Italy}

\begin{abstract}
We consider ionized hydrogen-like atoms accelerated by an external electric field to detect Unruh radiation. By applying quantum field theory in the Rindler spacetime, we show that the first-quantized description for hydrogen-like atoms cannot always be adopted. This is due to the frame-dependent definition of particles as positive and negative frequency field modes. We show how to suppress such a frame-dependent effect by constraining the atomic ionization and the electric field. We identify the physical regimes with nonvanishing atomic excitation probability due to the Unruh electromagnetic background. We recognize the observational limits for the Unruh effect via first-quantized atomic detectors, which appear to be compatible with current technology. Notably, the non-relativistic energy spectrum of the atom cannot induce coupling with the thermal radiation, even when special relativistic and general relativistic corrections are considered. On the contrary, the coupling with the Unruh radiation arises because of relativistic hyperfine splitting.
\end{abstract}
\maketitle

Accelerated detectors probe non-inertial quantum effects. Unruh and DeWitt~\cite{PhysRevD.14.870, PhysRevD.29.1047, hawking1980general} originally considered a single particle with acceleration $\alpha$ interacting with a scalar field via monopole coupling. In the non-inertial frame, the detector reveals a thermal background~\cite{PhysRevD.7.2850, Davies:1974th, PhysRevD.14.870} with temperature $T=\hbar \alpha/2\pi c k_{B}$, where $c$ is the speed of light and $k_{B}$ the Boltzmann constant. Similar descriptions have been used in more recent works to describe accelerated atoms as Unruh-DeWitt detectors \cite{PhysRevLett.91.243004, PhysRevA.74.023807, PhysRevLett.128.163603}.

More refined models including uniform external force as dynamical source for the acceleration superseded the original idealized description~\cite{PhysRevD.57.2403, PhysRevD.103.105023}. Furthermore, fully relativistic detectors have been considered \cite{PhysRevD.14.870, PARENTANI1995227}.

Here, we study accelerated atomic detectors by first principles within the relativistic Dirac theory in curved spacetime for hydrogen-like atoms. In the non-relativistic limit~\cite{falcone2022non, falcone2022}, we outline overlooked relativistic corrections and fundamental subtleties such as the frame-dependence of the particles number.

Despite the atom is initially prepared in the inertial laboratory frame as a non-relativistic bound state with a fixed number of electrons and nuclear particles, in the comoving accelerated frame, the energy and the number of the quantum particles are different~\cite{falcone2022}. A single electron appears as a superposition of states with varying energy and particles number and the electronic and nuclear structure is radically modified. The frame-dependent nature of particles -- at the origin of the Unruh effect~\cite{PhysRevD.7.2850} -- not only alters the background electromagnetic vacuum but also the electron and nuclear fields.

Such frame-dependence poses limits to adopting the familiar first-quantization description of the hydrogen-like atom in its proper frame and brings up difficulties in understanding light-matter interaction with noninertial observers. Lowering the acceleration suppresses the effect on the electrons and the other nuclear particles; however, this may also suppress the Unruh background electromagnetic vacuum, with the consequent decrease in the temperature. For a non-vanishing measurement of the Unruh effect, one needs an energy gap $\Delta E$ such that $\Delta E \lesssim k_{B} T$. Hence, the atomic spectrum must have a sufficiently fine structure to absorb the low-energy Unruh thermal photons.

Is it possible to suppress the frame-dependent effect on the electron while still detecting the electromagnetic thermal background?

We give a positive answer to the previous question by a rigorous analysis based on the quantum field theory in curved spacetime. Notwithstanding the suppressed frame-dependent effect for electrons, the hyperfine splitting provides the energy gap to reveal the Unruh radiation. We identify a specific parameter region in terms of the nuclear charge number $\mathcal{Z}$ and the electric field $E$ for the detection via first-quantized atomic detectors.

We assume that the atom is ionized with $1$ electron and $\mathcal{Z}>1$ protons. We consider an uniform electric field $\vec{E} = E \vec{e}_z$, with $E>0$, that produces an acceleration $\alpha$ along $\vec{e}_z$ such that $\alpha = (\mathcal{Z}-1) e E / M$, with $e$ the elementary charge and $M$ the atomic mass.

The electron is prepared in the laboratory frame as a non-relativistic particle. The inertial laboratory frame $(t,\vec{x})$ is defined by the Minkowski metric $\eta_{\mu\nu} = \text{diag} ( -c^2 , 1, 1, 1 )$. By using the interaction picture~\cite{falcone2022non}, we separate the free field theory from the interaction Lagrangian. The free electron field in the Minkowski frame $\hat{\psi}(t,\vec{x})$ reads
\begin{equation} \label{free_Dirac_field}
\hat{\psi}(t,\vec{x}) = \sum_{s=1}^2 \int_{\mathbb{R}^3} d^3 k \left[ u_s(\vec{k},t,\vec{x}) \hat{c}_s(\vec{k}) + v_s(\vec{k},t,\vec{x}) \hat{d}_s^\dagger(\vec{k}) \right],
\end{equation}
where $\hat{c}_s(\vec{k})$ and $\hat{d}_s(\vec{k})$ are annihilation operators for the particle and antiparticle with momentum $\vec{k}$ and spin number $s$. $u_s(\vec{k},t,\vec{x})$ and $v_s(\vec{k},t,\vec{x})$ are the positive and negative frequency modes such that
\begin{subequations}\label{free_Dirac_field_modes}
\begin{align}
& u_s(\vec{k},t,\vec{x}) = (2\pi)^{-3/2} e^{ -i\omega(\vec{k})t + i\vec{k} \cdot \vec{x} }  \tilde{u}_s(\vec{k}), \label{u} \\
& v_s(\vec{k},t,\vec{x}) = (2\pi)^{-3/2} e^{ i\omega(\vec{k})t - i\vec{k} \cdot \vec{x} }  \tilde{v}_s(\vec{k}), \label{v}
\end{align}
\end{subequations}
with $\omega (\vec{k}) = \sqrt{ ( m c^2/\hbar)^2 + c^2|\vec{k}|^2 }$ as the mode frequency. The spinors $\tilde{u}_s(\vec{k})$ and $\tilde{v}_s(\vec{k})$ are the orthonormal solutions of the Dirac equations in momentum space
\begin{subequations} \label{Dirac_uv_tilde}
\begin{align}
& \left[ \omega(\vec{k}) \gamma^0 - k_i \gamma^i - \frac{mc}{\hbar} \right] \tilde{u}_s(\vec{k}) = 0,\label{Dirac_u_tilde}\\
& \left[ \omega(\vec{k}) \gamma^0 - k_i \gamma^i + \frac{mc}{\hbar} \right] \tilde{v}_s(\vec{k}) = 0,\label{Dirac_v_tilde}\\
& \tilde{u}^\dagger_s(\vec{k}) \tilde{u}_{s'}(\vec{k}) = \tilde{v}^\dagger_s(\vec{k}) \tilde{v}_{s'}(\vec{k}) = \delta_{ss'},\\
& \tilde{u}^\dagger_s(\vec{k}) \tilde{v}_{s'}(-\vec{k}) = 0,
\end{align}
\end{subequations}
where $\gamma^\mu$ are the Dirac matrices~\cite{falcone2022non, falcone2022, falcone2023minkowski}.

The comoving accelerated frame $(T,\vec{X})$ is described by the Rindler metric $g_{\mu\nu}(T,\vec{X}) = \text{diag} ( -c^2 e^{2aZ}, 1, 1, e^{2aZ} )$, with $ a = \alpha/c^2$. We study the electron field by quantum field theory in Rindler spacetime (see Ref.~\cite{falcone2023minkowski} for the details).  The free electron field $\hat{\Psi}_\nu(T,\vec{X})$ is
\begin{align} \label{free_Dirac_field_Rindler}
\hat{\Psi}_\nu(T,\vec{X}) = & \sum_{s=1}^2 \int_0^\infty d\Omega \int_{\mathbb{R}^2} d^2 K_\perp  \nonumber \\
& \times \left[ U_{\nu s}(\Omega,\vec{K}_\perp,T,\vec{X}) \hat{C}_{\nu s}(\Omega,\vec{K}_\perp) \right. \nonumber \\
& \left. + V_{\nu s}(\Omega,\vec{K}_\perp,T,\vec{X}) \hat{D}_{\nu s}^\dagger(\Omega,\vec{K}_\perp) \right],
\end{align}
where $\hat{C}_{\nu s}(\Omega,\vec{K}_\perp)$ and $\hat{D}_{\nu s}(\Omega,\vec{K}_\perp)$ annihilate the electron and the positron of the $\nu$ wedge with spin number $s$, frequency $\Omega$ and transverse momentum $\vec{K}_\perp$ represented by the positive and negative frequency modes
\begin{subequations}\label{UV_UV_tilde}
\begin{align}
& U_{\nu s}(\Omega,\vec{K}_\perp,T,\vec{X}) = e^{i \vec{K}_\perp \cdot \vec{X}_\perp - i \Omega T} \tilde{W}_{\nu s}(\Omega,\vec{K}_\perp,Z),\label{U_U_tilde} \\
& V_{\nu s}(\Omega,\vec{K}_\perp,T,\vec{X}) = e^{-i \vec{K}_\perp \cdot \vec{X}_\perp + i \Omega T} \tilde{W}_{\nu s}(-\Omega,-\vec{K}_\perp,Z),\label{V_V_tilde}
\end{align}
\end{subequations}
with
\begin{align}\label{W_tilde_W_tilde}
& \tilde{W}_{\nu s'}(\Omega,\vec{K}_\perp,Z_\nu(z)) = \frac{1}{2 \pi^2} \sqrt{\frac{\kappa (\vec{K}_\perp)}{c a} \cosh \left( \frac{\beta}{2} \Omega \right)}   \nonumber \\
 &\times \sum_{\sigma=\pm} K_{\sigma s_\nu i \Omega / c a - 1/2} \left( \kappa (\vec{K}_\perp) \frac{e^{s_\nu aZ}}{a} \right)  \left[ \frac{- s_\nu i c} {\kappa (\vec{K}_\perp)} \right.\nonumber \\
 &\times \left. \gamma^0 \left( K_1 \gamma^1 + K_2 \gamma^2 + \frac{m c}{\hbar} \right) \right]^{(1 - \sigma)/2}  \tilde{\mathfrak{W}}_{\nu s}(\Omega, \vec{K}_\perp) ,
\end{align}
and where $s_\nu$ is the sign of the wedge (i.e., $s_\text{L}=-1$ and $s_\text{R}=1$), $\kappa (\vec{K}_\perp) =  \sqrt{ (m c/\hbar)^2 + |\vec{K}_\perp|^2 }$ is the reduced momentum, $K_\zeta (\xi)$ is the modified Bessel function of the second kind and $\tilde{\mathfrak{W}}_{\nu s}(\Omega, \vec{K}_\perp)$ are orthonormal bases for the eigenspace of $c \gamma^0 \gamma^3$ with eigenvalue $1$, i.e.,
\begin{subequations} \label{W_tilde_constraints}
\begin{align}
& c\gamma^0 \gamma^3 \tilde{\mathfrak{W}}_{\nu s}(\Omega, \vec{K}_\perp) =  \tilde{\mathfrak{W}}_{\nu s}(\Omega, \vec{K}_\perp),\label{W_tilde_constraints_b} \\
&  \tilde{\mathfrak{W}}^\dagger_{\nu s} (\Omega,\vec{K}_\perp)  \tilde{\mathfrak{W}}_{\nu s'} (\Omega,\vec{K}_\perp) = \delta_{ss'} \label{W_tilde_constraints_c}.
\end{align}
\end{subequations}

In Ref.~\cite{falcone2023minkowski}, we reported the following Bogoliubov transformation relating the Minkowski particle creator $\hat{c}_s(\vec{k})$ to Rindler operators 
\begin{align}\label{Bogoliubov_transformations_3_Rindler_4_a}
& \hat{c}_s(\vec{k}) = \sum_{\nu=\{\text{L},\text{R}\}}  \sum_{s'=1}^2 \int_\mathbb{R} d\Omega \int_{\mathbb{R}^2} d^2 K_\perp   \nonumber \\
& \times  \alpha_{\nu}(\vec{k},\Omega,\vec{K}_\perp) \tilde{u}^\dagger_s(\vec{k})  \tilde{\mathfrak{W}}_{\nu s'}(\Omega, \vec{K}_\perp) \nonumber \\
& \times \left[ \theta(\Omega) \hat{C}_{\nu s'}(\Omega,\vec{K}_\perp)+  \theta(-\Omega)\hat{D}_{\nu s'}^\dagger(-\Omega,-\vec{K}_\perp) \right],
\end{align}
with
\begin{align}\label{Bogoliubov_coefficient}
& \alpha_{\nu}(\vec{k},\Omega,\vec{K}_\perp) = \frac{1}{\pi} \delta^2(\vec{k}_\perp - \vec{K}_\perp) \sqrt{\frac{\kappa (\vec{k}_\perp)}{2 \pi c a} \cosh \left( \frac{\pi \Omega}{c a} \right)}  \nonumber \\
& \times \sum_{\sigma = \pm} \left[ s_\nu i \frac{\omega (\vec{k}) + c k_3}{c \kappa (\vec{k}_\perp)} \right]^{(\sigma - 1)/2} \int_{\mathbb{R}} dz \theta(s_\nu z) e^{ - ik_3 z } \nonumber \\
& \times K_{\sigma s_\nu i \Omega / c a - 1/2} ( s_\nu \kappa (\vec{K}_\perp)  z ),
\end{align}
and we showed the following representation of the Minkowski vacuum in the Rindler spacetime
\begin{align}\label{0_M_0_LR}
|0_\text{M} \rangle \propto & \exp \left( - i \sum_{\nu=\{\text{L},\text{R}\}} s_\nu \sum_{s=1}^2 \sum_{s'=1}^2 \int_0^{+\infty} d\Omega \int_{\mathbb{R}^2} d^2 K_\perp   \right. \nonumber \\ 
& \times e^{-\pi \Omega / c a}  \tilde{\mathfrak{W}}^\dagger_{\nu s}(\Omega, \vec{K}_\perp)  \tilde{\mathfrak{W}}_{\bar{\nu} s'}(-\Omega, \vec{K}_\perp)  \nonumber \\ 
& \left. \times  \hat{C}^\dagger_{\nu s}(\Omega,\vec{K}_\perp)\hat{D}^\dagger_{\bar{\nu} s'}(\Omega,-\vec{K}_\perp) \right) |0_\text{L},0_\text{R} \rangle,
\end{align}
where $|0_\text{M} \rangle$ is the Minkowski vacuum, $\bar{\nu}$ is the opposite of $\nu$ (i.e., $\bar{\nu}=\text{L} $ if $\nu= \text{R}$ and $\bar{\nu}= \text{R}$ if $\nu= \text{L}$) and $|0_\text{L},0_\text{R} \rangle$ is the Rindler vacuum state. By using Eqs.~(\ref{Bogoliubov_transformations_3_Rindler_4_a}) and (\ref{0_M_0_LR}), one can see that any non-relativistic single-electron prepared in the inertial frame appears as a superposition of states with varying energy and particle number in the accelerated frame.

In the case of scalar fields, such frame-dependent effect is suppressed when the acceleration $\alpha$ is sufficiently low and the particle state is localized in the approximately Minkowskian region of the Rindler spacetime (i.e., where $g_{\mu\nu}(T,\vec{X})\approx \eta_{\mu\nu}$)~\cite{falcone2022}. Any nonrelativistic Minkowski single particle appears as a nonrelativistic Rindler particle in the accelerated frame if $\alpha$ is such that

\begin{equation}\label{low_acceleration}
\frac{\hbar a}{mc} \lesssim \epsilon^{3/2}
\end{equation}
and the localization in $\vec{x}$ in such that
\begin{equation}\label{local}
|a z - 1| \lesssim \epsilon,
\end{equation}
where $\epsilon=\hbar \Omega/mc^2$ is the nonrelativistic parameter defined as the ratio between the nonrelativistic energy $\hbar \Omega$ and the mass energy $mc^2$. The resulting Rindler single-particle is created over the Unruh background $|0_\text{M} \rangle$, which is in a superposition of Rindler particles. These background particles are mostly localized far from the region (\ref{local}) and close to the Rindler horizon. Hence, they can be ignored for the local detection of the Unruh effect.

When $\hbar a / mc \lesssim \epsilon^{3/2}$, $|az-1| \lesssim \epsilon$ and $|\Omega| / c a \lesssim 1$, the Bessel functions appearing in Eq.~(\ref{Bogoliubov_coefficient}) are approximated as~\cite{olver2014asymptotics}
\begin{equation}\label{Bessel_limits_a}
K_{ \pm i \Omega / c a - 1/2} ( \kappa (\vec{K}_\perp)  z ) \approx 0,
\end{equation}
and, when $\hbar a / mc \lesssim \epsilon^{3/2}$, $\hbar |\vec{K}_\perp|/mc \lesssim \epsilon^{1/2}$, $|az-1| \lesssim \epsilon$ and $|\Omega| / c a \gg 1$, as
\begin{align}\label{Bessel_limits_b}
& \sqrt{ \cosh \left( \frac{\pi \Omega}{c a} \right) } K_{ \pm i \Omega / c a - 1/2} ( \kappa (\vec{K}_\perp)  z ) \approx \pi \left(\frac{\hbar a}{\sqrt{2} m c}\right)^{1/3}\nonumber \\
 &  \times \text{Ai} \left( \left(\frac{\sqrt{2} m c}{\hbar a}\right)^{2/3} \left(\left(\frac{\hbar |\vec{K}_\perp|}{\sqrt{2} m c}\right)^{2} + a z  - \frac{\hbar \Omega}{m c^2}  \right) \right),
\end{align}
where $\text{Ai}(\xi)$ is the Airy function. We use Eqs.~(\ref{Bessel_limits_a}) and (\ref{Bessel_limits_b}) in Ref.~\cite{Supplementary} to prove that the electron is seen as a nonrelativistic single particle in both frames if the acceleration is constrained by Eq.~(\ref{low_acceleration}) and the electron is localized in the region given by Eq.~(\ref{local}).

Now we discuss the interaction between the electron and the classic electromagnetic field. The particle is affected by the potential energy $V_\text{ext}$ due to the external electric field $\vec{E}$ and the potential energy $V_\text{nuc}$ due to the nuclear Coulomb interaction. In the comoving frame, $V_\text{ext}$ is \cite{Supplementary}
\begin{equation}\label{V_ext}
V_\text{ext}(Z) = \frac{1-e^{-2 a Z}}{2} \frac{e E}{a}.
\end{equation}
The nuclear potential energy $V_\text{nuc}$ is
\begin{equation}
V_\text{nuc}(R) = - \epsilon_\text{QED}^{1/2} \frac{\hbar c}{R},
\end{equation}
where $R=|\vec{X}|$ is the radial coordinate, $\epsilon_\text{QED} = (\mathcal{Z} \alpha_0)^2$ is the quantum electrodynamics (QED) coupling and $\alpha_0$ the fine-structure constant.

The electron is pulled away from its orbit by $V_\text{ext}$ while it is dragged by the accelerating nucleus via $V_\text{nuc}$. If $E$ is sufficiently large, the electron escapes from the nuclear Coulomb barrier via quantum tunneling, compromising the atomic stability. To avoid complete ionization, we require a small $E$ such that
\begin{equation}\label{E_sufficiently_small}
|V_\text{ext}(R_0)| \ll |E^{(0)}_0|,
\end{equation}
with $E^{(0)}_0 = -\epsilon_\text{QED} \mu c^2/2 $ as the ground state of $V_\text{nuc}$, $\mu=(m + M_\text{N})/m M_\text{N} \approx m$ as the reduced mass, $M_\text{N} \approx M$ as the nuclear mass, $R_0=a_0/\mathcal{Z}$ as the atomic radius and $a_0 = \hbar/m c \alpha_0$ as the Bohr radius. Hence, we assume that the external force $V_\text{ext}$ simply perturbs the spectrum of $V_\text{nuc}$ via Stark effect.

Equation (\ref{E_sufficiently_small}) reads
\begin{equation}\label{a_sufficiently_small_local}
a R_0 \ll \epsilon_\text{QED} \frac{(\mathcal{Z}-1)m}{2 M},
\end{equation}
or, equivalently,
\begin{equation} \label{Z_E_constraint_1_3}
E \ll \frac{(\mathcal{Z} \alpha_0)^3}{2} \frac{m^2 c^3}{\hbar e}.
\end{equation}
Notice that the electron is localized inside the region $R\lesssim R_0$ since $V_\text{nuc}$ dominates over $V_\text{ext}$. Notice also that $\epsilon_\text{QED} \ll 1$ and $(\mathcal{Z}-1) m \ll M$. Hence, from Eqs.~(\ref{V_ext}) and (\ref{a_sufficiently_small_local}), one concludes that the electron is localized where the electric field is approximately uniform.

Equation (\ref{E_sufficiently_small}) guarantees a lifetime $\tau$ for the atom that is exponentially increasing for decreasing electric field. Indeed, by using the WKB approximation, one can find the following ionization rate \cite{landau1991quantum}
\begin{equation}\label{tau}
\frac{1}{\tau} \approx  \frac{16}{\hbar R_0 e E}  \left( E^{(0)}_0 \right)^2 \exp \left( \frac{4 E^{(0)}_0}{3 R_0 e E} \right).
\end{equation}

Notice that $\epsilon_\text{QED} mc^2$ is the order of the nonrelativistic atomic energies. Indeed, the spectrum of $V_\text{nuc}$ is
\begin{equation}\label{atom_spectrum}
E^{(0)}_n = - \frac{\epsilon_\text{QED}}{2 (n+1)^2} \mu c^2.
\end{equation}
By comparing Eq.~(\ref{a_sufficiently_small_local}) with Eq.~(\ref{local}), one finds out that the localization condition (\ref{local}) is already met by configurations that satisfy Eq.~(\ref{a_sufficiently_small_local}). Furthermore, Eq.~(\ref{a_sufficiently_small_local}) leads to
\begin{equation}\label{a_sufficiently_small}
\frac{\hbar a}{mc} \ll \epsilon^{3/2}_\text{QED} \frac{(\mathcal{Z}-1)m}{2M},
\end{equation}
which is a sufficient condition for Eq.~(\ref{low_acceleration}). By constraining $E$ and $\mathcal{Z}$ accordingly to Eq.~(\ref{Z_E_constraint_1_3}), one guarantees the atom stability and the first-quantization electron description in the accelerated frame. The atom does not ionize and the electron appears as a nonrelativistic single-particle bound state in both frames.

We next consider the interaction between the accelerated atom and the electromagnetic Unruh background \cite{PhysRevD.14.870, PhysRevD.29.1047, PhysRevA.74.023807}. The electron is excited by absorbing a photon with the energy $\Delta E_n = E_n - E_0$ of the $n$-th electronic transition. The event is detectable if
\begin{equation}\label{Delta_E_n_order_a}
\Delta E_n \lesssim k_B T,
\end{equation}
and if the atom has a sufficiently large lifetime such that
\begin{equation}\label{upper_limit_tau}
\tau \gtrsim \frac{\hbar}{\Delta E_n}.
\end{equation}

Equation (\ref{Delta_E_n_order_a}) guarantees a nonvanishing probability for the electron to interact with photons described by the following Boltzmann distribution
\begin{equation}\label{Boltzmann_distribution}
P_\text{B} = \frac{1}{e^{\Delta E_n/k_\text{B} T} - 1}.
\end{equation}
A more refined constant than Eq.~(\ref{Delta_E_n_order_a}) can be imposed by assuming a lower bound for $P_\text{B}$, i.e.,
\begin{equation}\label{lower_limit}
P_\text{B} < P_\text{min},
\end{equation}
with $P_\text{min}<1$. Equation (\ref{upper_limit_tau}), instead, ensures that the absorption spectrum of the atom is narrow around $\Delta E_n$. Given the exponential growth of the atom lifetime for smaller $E$ [see Eq.~(\ref{tau})], it is safe to assume that Eq.~(\ref{upper_limit_tau}) gives an almost exact lower limit for $\tau$, i.e.,
\begin{equation}\label{upper_limit_tau_2}
\tau > \frac{\hbar}{\Delta E_n}.
\end{equation}

The states and energies of the spectrum $E_n$ are the solutions of the Dirac equation in Rindler spacetime for hydrogen-like atoms with the interaction potential $V_\text{ext}$. They can be computed perturbatively by considering the non-relativistic hydrogen-like spectrum $E^{(0)}_n$ [see Eq.~(\ref{atom_spectrum})] perturbed by $V_\text{ext}$ and by relativistic corrections coming from the Rindler-Dirac equation.

The energies gaps of the unperturbed Hamiltonian $\Delta E^{(0)}_n = E^{(0)}_n - E^{(0)}_0$ in Eq.~(\ref{atom_spectrum}) are of the order
\begin{equation}\label{Delta_E_0_n_order}
\Delta E^{(0)}_n \sim \epsilon_\text{QED} mc^2.
\end{equation}
By plugging Eq.~(\ref{Delta_E_0_n_order}) in Eq.~(\ref{Delta_E_n_order_a}), one finds that the lower bound for the electric field is
\begin{equation}
E \gtrsim \frac{2\pi (\mathcal{Z} \alpha_0)^2}{\mathcal{Z}-1} \frac{m M c^3}{ \hbar e},
\end{equation}
which is way larger than the upper bound (\ref{Z_E_constraint_1_3}). Hence, $\Delta E^{(0)}_n$ does not induce coupling with the electromagnetic background for any stable configuration.

Perturbations of $E^{(0)}_n$ do not significantly change the energies gaps, unless they break the spin degeneracy of the atomic ground state. In that case, the first level $E^{(0)}_0$ splits into the actual ground state $E_0$ and the first excited state $E_1$, with $\Delta E = E_1 - E_0 \ll \epsilon_\text{QED} mc^2$.

In Ref.~\cite{Supplementary}, we show that the Rindler-Dirac equation for the hydrogen-like atom with potentials $V_\text{nuc}$ and $V_\text{ext}$ have a degenerate minimum energy level. Hence, the external electron field $V_\text{ext}$ and the special and general relativity corrections do not break the spin degeneracy of $E^{(0)}_0$.

One has to look at the hyperfine structure to see a split of $E^{(0)}_0$ due to quantum electrodynamics corrections. The electron-nucleus interaction via spin-spin coupling generates the following energy gap \cite{bethe2013quantum}
\begin{equation}\label{Delta_E_hf}
\Delta E_\text{hf} = \begin{cases}
\frac{1}{3 \pi} (2I + 1) Z^3 \alpha_0^4 g \frac{m^2c^2}{M_\text{P}} & \text{if } I \neq 0 \\
0 & \text{if } I = 0
\end{cases},
\end{equation}
with $M_\text{P}$ as the proton mass. $I$ is the quantum number such that $|\vec{I}|^2 = I(I+1)$, where $\vec{I}$ is the nucleus spin. $g$ is the effective $g$-factor defined as follows: $\vec{\mu} = (g \hbar e/2M_\text{P}) \vec{I}$, where $\vec{\mu}$ is the magnetic moment of the nucleus resulting from its spin.

Notice that the selection rule that forbids transitions between levels with vanishing azimuthal quantum number $\ell = 0$ breaks down due to the Stark effect. Hence, the absorption of photons coupled to the hyperfine structure is allowed.

By plugging Eq.~(\ref{Delta_E_hf}) in Eq.~(\ref{lower_limit}), one finds that the atomic hyperfine structure produces a measurable Boltzmann distribution when $I \neq 0$ and when
\begin{equation}\label{hf_measurable_Unruh}
\left[ \exp \left( \frac{2 (2I + 1) \mathcal{Z}^3 \alpha_0^4 g}{3(\mathcal{Z}-1)} \frac{M m^2 c^3}{M_\text{P} \hbar e E} \right) - 1 \right]^{-1} < P_\text{min} .
\end{equation}
Furthermore, the atom has a sufficiently long lifetime when [see Eqs.~(\ref{upper_limit_tau_2}) and (\ref{Delta_E_hf})]
\begin{equation}\label{upper_limit_hf}
\frac{ \hbar e E}{m^2 c^3} \exp \left( \frac{2 (\mathcal{Z} \alpha_0)^3 }{3} \frac{ m^2 c^3}{\hbar e E} \right) > \frac{12 \pi \mathcal{Z}^2 \alpha_0}{(2I + 1) g} \frac{M_\text{P}}{m} .
\end{equation}

%%%%%%%%%%%%%%%%%%%%%%%%%%%%%%%%%%%%%%%%%%%%%%%%%%%%%%%%%%%%%%%%%%%%%%%%%%%%%%%%%%%%%%%%%%
\begin{figure}
\includegraphics{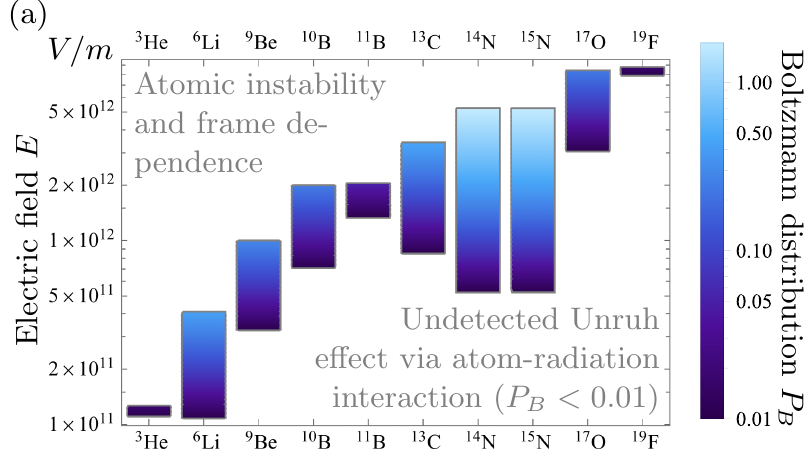}
\includegraphics{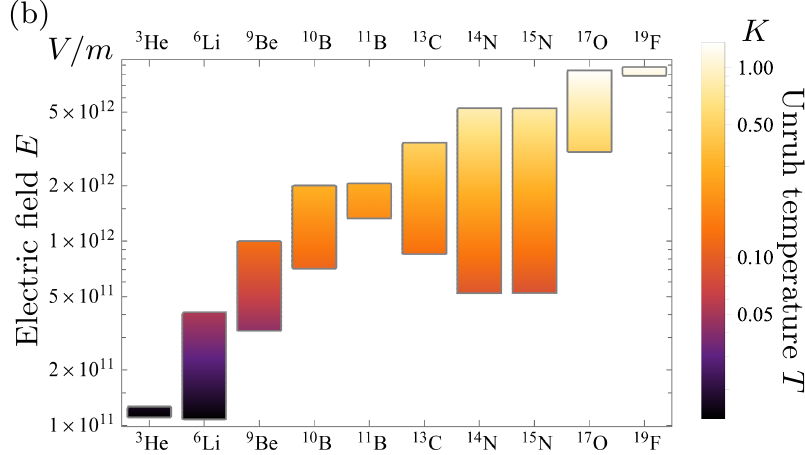}
\caption{Observation window for the Unruh effect via first-quantized atomic detectors. The constrained variable is the accelerating electric field $E$ for each nuclear configuration $\mathcal{Z}$, $M$, $I$, $g$ \cite{STONE200575}. The upper limit for $E$ [see Eq.~(\ref{upper_limit_hf})] guarantees the stability of the atomic bound state and the first-quantized description of the electron in the accelerated frame. Above this limit, the electron escapes from the Coulomb potential via tunneling and it appears as a superposition of states with different energies and number of particles in the accelerated frame. The lower limit [see Eq.~(\ref{hf_measurable_Unruh})], instead, ensures the detection of the Unruh effect from the electromagnetic thermal background via light-matter interaction. Below such limit, the hyperfine structure of the atom produces an energies gap that is too large for the Boltzmann distribution to be detected. In (a) we show the Boltzmann distribution for the atom-radiation interaction [see Eq.~(\ref{Boltzmann_distribution_hp})]. In (b), we show the Unruh temperature $T$ of the electromagnetic background in the accelerated frame for different configurations [see Eq.~(\ref{T_E_Z})].} \label{Fig1}
\end{figure}
%%%%%%%%%%%%%%%%%%%%%%%%%%%%%%%%%%%%%

Equations (\ref{hf_measurable_Unruh}) and (\ref{upper_limit_hf}) define the regime of parameters $E$, $\mathcal{Z}$, $M$, $I$, $g$ for the detection of the Unruh effect via first-quantized atomic detectors. The results are shown in Fig.~\ref{Fig1}, where for some nuclear configurations we plot the range of validity for the electric field $E$. In Fig.~\ref{Fig1}, we also plot the Boltzmann distribution
\begin{equation}\label{Boltzmann_distribution_hp}
P_\text{B} = \left[ \exp \left( \frac{(2I + 1) \mathcal{Z}^3 \alpha_0^4 g}{3(\mathcal{Z}-1)} \frac{M m^2 c^3}{M_\text{P} \hbar e E} \right) - 1 \right]^{-1}
\end{equation}
and the Unruh temperature
\begin{equation}\label{T_E_Z}
 T = \frac{\mathcal{Z} - 1 }{2 \pi } \frac{ \hbar e E }{k_\text{B} M c}.
\end{equation}
To overcome background noises, a sufficiently large $T$ is needed and requires high ionization $\mathcal{Z}$.

Various experimental proposals have been reported to test the Unruh effect. The proposals include the depolarization of electrons in storage rings \cite{BELL1983131, BELL1987488}, Penning traps \cite{Rogers:1988zz}, ultraintense lasers \cite{PhysRevLett.83.256, Schutzhold}. The growing interest is motivated by the ever-improving experimental equipment that allows to reach high accelerations \cite{Brodin_2008}. Besides electrons, also uniformly accelerated protons have been considered as Unruh-DeWitt detectors via acceleration-induced weak-interaction decay \cite{PhysRevD.56.953, PhysRevLett.87.151301, PhysRevD.67.065002} and photon emission \cite{PhysRevD.46.3450, Higuchi:1992we}.

In this manuscript, we analyzed the electron in the accelerated atom by the non-relativistic limit of a Dirac field in Rindler spacetime \cite{falcone2022, falcone2022non, falcone2023minkowski}. While considering hyperfine splitting, we addressed three problems: (i) the instability of the atom due to a strong accelerating field; (ii) the frame-dependent nature of the electron; (iii) the detectability of the Unruh effect due to the electromagnetic radiation.
We have shown that (i) and (ii) impose an upper boundary condition for the electric field accelerating the ionized hydrogen-like atom. Taking into account (iii), we determined an observation window in the $E$ and $\mathcal{Z}$ plane [see Fig.~\ref{Fig1}]. Surprisingly, quantitative estimates unveil that the effect can be detected with electric fields within the reach of the modern technologies of high-power lasers and nuclear magnetic resonance.

\bibliography{bibliography}

\end{document}

\title{Supplemental materials}

\maketitle

\section{Suppressing the frame dependent effect for Dirac single particles}

In this section, we adapt the discussion of Ref.~\cite{falcone2022} to the case of Dirac fields. We consider a single fermionic particle $| \psi \rangle $ prepared by an inertial observer and we derive its representation in the Rindler-Fock space by using the Bogoliubov transformation (\ref{Bogoliubov_transformations_3_Rindler_4_a}). We assume that the acceleration is constrained by Eq.~(\ref{low_acceleration}) and that the particle is nonrelativistic and localized in (\ref{local}). We show that the Rindler-Fock representative for $| \psi \rangle $ is approximated by a nonrelativistic Rindler single particle created over the Minkowski vacuum $|0_\text{M} \rangle$, which, in turns, is mainly populated by Rindler particles and antiparticles localized far from (\ref{local}).

We start from the general definition of Minkowski single particles $| \psi \rangle$, with wave function $\psi(x)$. The state is defined as follows
\begin{equation}\label{single_particle}
| \psi \rangle = \hat{\mathfrak{C}}^\dagger_\psi |0_\text{M}\rangle,
\end{equation}
with
\begin{equation}\label{C_psi}
\hat{\mathfrak{C}}_\psi = \sum_{s=1}^2  \int_{\mathbb{R}^3} d^3 k \tilde{\psi}^*_s(\vec{k}) \hat{c}_s(\vec{k})
\end{equation}
as the particle annihilator and with $\tilde{\psi}_s(\vec{k})$ as the wave function in the spin-momentum representation, i.e.,
\begin{equation}\label{psi_psi_tilde}
\psi(\vec{x}) = \sum_{s=1}^2 \int_{\mathbb{R}^3} d^3 k \tilde{\psi}_s(\vec{k}) u_s(\vec{k},0,\vec{x}).
\end{equation}
The nonrelativistic condition for the particle is
\begin{align}\label{nonrelativistic_wavefunction}
& \tilde{\psi}_s(\vec{k}) \approx 0 && \text{if} && \frac{\hbar |\vec{k}|}{mc} \gg \epsilon^{1/2}.
\end{align}
The localization condition, instead, is
\begin{align}\label{local_wavefunction}
& \psi(\vec{x}) \approx 0 && \text{if} && |a z - 1 | \gg \epsilon.
\end{align}

By using the Bogoliubov transformation (\ref{Bogoliubov_transformations_3_Rindler_4_a}), we write $\hat{\mathfrak{C}}^\dagger_\psi$ in terms of Rindler operators,
\begin{align}\label{C_Bogoliubov_transformations}
& \hat{\mathfrak{C}}_\psi = \sum_{\nu=\{\text{L},\text{R}\}} \sum_{s=1}^2 \sum_{s'=1}^2  \int_{\mathbb{R}^3} d^3 k \int_\mathbb{R} d\Omega \int_{\mathbb{R}^2} d^2 K_\perp  \tilde{\psi}^*_s(\vec{k})  \nonumber \\
& \times  \alpha_{\nu}(\vec{k},\Omega,\vec{K}_\perp) \tilde{u}^\dagger_s(\vec{k})  \tilde{\mathfrak{W}}_{\nu s'}(\Omega, \vec{K}_\perp) \nonumber \\
& \times \left[ \theta(\Omega) \hat{C}_{\nu s'}(\Omega,\vec{K}_\perp)+  \theta(-\Omega)\hat{D}_{\nu s'}^\dagger(-\Omega,-\vec{K}_\perp) \right].
\end{align}
As a consequence of the Dirac delta function appearing in the explicit form of $\alpha_{\nu}(\vec{k},\Omega,\vec{K}_\perp)$ [see Eq.~(\ref{Bogoliubov_coefficient})] and the nonrelativistic condition (\ref{nonrelativistic_wavefunction}), $\hat{\mathfrak{C}}^\dagger_\psi$ creates Rindler particles and destroys Rindler antiparticles with nonrelativistic transverse momentum, i.e., $\hbar |\vec{K}_\perp|/mc \lesssim \epsilon^{1/2}$. Furthermore, Eqs.~(\ref{Bogoliubov_coefficient}) and (\ref{nonrelativistic_wavefunction}) lead to the following approximation
\begin{align}\label{C_Bogoliubov_transformations_2}
& \hat{\mathfrak{C}}_\psi \approx \sum_{\nu=\{\text{L},\text{R}\}} \sum_{s=1}^2 \sum_{s'=1}^2  \int_{\mathbb{R}^3} d^3 k \int_\mathbb{R} d\Omega \int_{\mathbb{R}^2} d^2 K_\perp  \tilde{\psi}^*_s(\vec{k})  \nonumber \\
& \times  \beta_{\nu}(\vec{k},\Omega,\vec{K}_\perp) \tilde{u}^\dagger_s(\vec{k})  \tilde{\mathfrak{W}}_{\nu s'}(\Omega, \vec{K}_\perp) \nonumber \\
& \times \left[ \theta(\Omega) \hat{C}_{\nu s'}(\Omega,\vec{K}_\perp)+  \theta(-\Omega)\hat{D}_{\nu s'}^\dagger(-\Omega,-\vec{K}_\perp) \right],
\end{align}
with
\begin{align}\label{beta}
& \beta_{\nu}(\vec{k},\Omega,\vec{K}_\perp) = \frac{1}{\pi} \delta^2(\vec{k}_\perp - \vec{K}_\perp) \sqrt{\frac{m}{2 \pi \hbar  a} \cosh \left( \frac{\pi \Omega}{c a} \right)}  \nonumber \\
& \times \sum_{\sigma = \pm} ( s_\nu i )^{(\sigma - 1)/2}  \int_{\mathbb{R}} dz \theta(s_\nu z) e^{ - ik_3 z } \nonumber \\
& \times K_{\sigma s_\nu i \Omega / c a - 1/2} ( s_\nu \kappa (\vec{K}_\perp)  z ).
\end{align}

By recalling Eq.~(\ref{u}), we write Eq.~(\ref{C_Bogoliubov_transformations_2}) in the following way
\begin{align}\label{C_Bogoliubov_transformations_3}
& \hat{\mathfrak{C}}_\psi \approx \sum_{\nu=\{\text{L},\text{R}\}} \sum_{s=1}^2  \int_\mathbb{R} d\Omega \int_{\mathbb{R}^2} d^2 K_\perp  \chi_{\nu s}(\Omega,\vec{K}_\perp) \nonumber \\
& \times \left[ \theta(\Omega) \hat{C}_{\nu s}(\Omega,\vec{K}_\perp)+  \theta(-\Omega)\hat{D}_{\nu s}^\dagger(-\Omega,-\vec{K}_\perp) \right],
\end{align}
with
\begin{align}\label{chi}
& \chi_{\nu s}(\Omega,\vec{K}_\perp) = \frac{1}{2 \pi^2} \sqrt{\frac{m}{\hbar a} \cosh \left( \frac{\pi \Omega}{c a} \right)}  \int_{\mathbb{R}^3} d^3 x \theta(s_\nu z) \nonumber \\
& \times \sum_{\sigma = \pm}  ( s_\nu i )^{(\sigma - 1)/2} K_{\sigma s_\nu i \Omega / c a - 1/2} ( s_\nu \kappa (\vec{K}_\perp)  z ) \nonumber \\
& \times  \sum_{s'=1}^2 \int_{\mathbb{R}^3} d^3 k \tilde{\psi}^*_{s'}(\vec{k}) u^\dagger_{s'}(\vec{k},0,\vec{x})  \tilde{\mathfrak{W}}_{\nu s}(\Omega, \vec{K}_\perp).
\end{align}
By using Eq.~(\ref{psi_psi_tilde}), Eq.~(\ref{chi}) reads as
\begin{align}\label{chi_2}
& \chi_{\nu s}(\Omega,\vec{K}_\perp) = \frac{1}{2 \pi^2} \sqrt{\frac{ m}{\hbar a} \cosh \left( \frac{\pi \Omega}{c a} \right)}  \int_{\mathbb{R}^3} d^3 x \theta(s_\nu z) \nonumber \\
& \times \sum_{\sigma = \pm}  ( s_\nu i )^{(\sigma - 1)/2} K_{\sigma s_\nu i \Omega / c a - 1/2} ( s_\nu \kappa (\vec{K}_\perp)  z ) \nonumber \\
& \times  \psi^\dagger(\vec{x}) \tilde{\mathfrak{W}}_{\nu s}(\Omega, \vec{K}_\perp).
\end{align}
The localization condition (\ref{local_wavefunction}) can be used in Eq.~(\ref{chi_2}) to obtain the following approximation 
\begin{subequations}\label{chi_3}
\begin{align}
& \chi_{\text{R} s}(\Omega,\vec{K}_\perp) \approx  \frac{1}{2 \pi^2} \sqrt{\frac{ m}{\hbar a} \cosh \left( \frac{\pi \Omega}{c a} \right)}  \int_{\mathbb{R}^3} d^3 x\nonumber \\
& \times  \theta(\epsilon-|az-1| ) \sum_{\sigma = \pm}  i^{(\sigma - 1)/2} \nonumber \\
& \times K_{\sigma  i \Omega / c a - 1/2}(  \kappa (\vec{K}_\perp)  z )  \psi^\dagger(\vec{x}) \tilde{\mathfrak{W}}_{\text{R} s}(\Omega, \vec{K}_\perp), \label{chi_3_a}\\
& \chi_{\text{L} s}(\Omega,\vec{K}_\perp) \approx 0,
\end{align}
\end{subequations}
which means that $\hat{\mathfrak{C}}^\dagger_\psi$ creates Rindler particles and destroys Rindler antiparticles in the right wedge only. By using again Eq.~(\ref{psi_psi_tilde}), Eq.~(\ref{chi_3_a}) reads as
\begin{align}\label{chi_4}
& \chi_{\text{R} s}(\Omega,\vec{K}_\perp) \approx \frac{1}{2 \pi^2} \sqrt{\frac{ m}{\hbar a} \cosh \left( \frac{\pi \Omega}{c a} \right)}  \sum_{s'=1}^2 \int_{\mathbb{R}^3} d^3 k  \nonumber \\
& \times  \tilde{\psi}^*_{s'}(\vec{k})  \sum_{\sigma = \pm}   i^{(\sigma - 1)/2} \int_{\mathbb{R}^3} d^3 x \theta(\epsilon-|az-1| )  \nonumber \\
& \times  u^\dagger_{s'}(\vec{k},0,\vec{x})  \tilde{\mathfrak{W}}_{\text{R} s}(\Omega, \vec{K}_\perp) K_{\sigma  i \Omega / c a - 1/2} (  \kappa (\vec{K}_\perp)  z ).
\end{align}

We now consider the limit of low acceleration (\ref{low_acceleration}) and we use Eqs.~(\ref{Bessel_limits_a}) and (\ref{Bessel_limits_b}) in Eq.~(\ref{chi_4}). By using Eq.~(\ref{Bessel_limits_a}), we find that $\chi_{\text{R} s}(\Omega,\vec{K}_\perp)$ is vanishing when $|\Omega| / c a \lesssim 1$, and hence $\hat{\mathfrak{C}}^\dagger_\psi$ creates Rindler particles and destroys Rindler antiparticles with frequency $|\Omega| \gg c a $. By using Eq.~(\ref{Bessel_limits_b}), one can write Eq.~(\ref{chi_4}) in the limit $|\Omega| \gg c a $ as follows
\begin{align}\label{chi_5}
& \chi_{\text{R} s}(\Omega,\vec{K}_\perp) \approx \frac{1-i}{2 \pi} \sqrt{\frac{m}{\hbar a}}   \left(\frac{\hbar a}{\sqrt{2} m c}\right)^{1/3} \sum_{s'=1}^2 \int_{\mathbb{R}^3} d^3 k\nonumber \\
& \times  \tilde{\psi}^*_{s'}(\vec{k}) \int_{\mathbb{R}^3} d^3 x \theta(\epsilon-|az-1| )  u^\dagger_{s'}(\vec{k},0,\vec{x})  \tilde{\mathfrak{W}}_{\text{R} s}(\Omega, \vec{K}_\perp)\nonumber \\
 &  \times \text{Ai} \left( \left(\frac{\sqrt{2} m c}{\hbar a}\right)^{2/3} \left(\left(\frac{\hbar |\vec{K}_\perp|}{\sqrt{2} m c}\right)^{2} + a z  - \frac{\hbar \Omega}{m c^2}  \right) \right).
\end{align}
Owing to Eq.~(\ref{u}), Eq.~(\ref{chi_5}) reads as
\begin{align}\label{chi_6}
& \chi_{\text{R} s}(\Omega,\vec{K}_\perp) \approx \frac{1-i}{(2 \pi)^{5/3}} \sqrt{\frac{m}{\hbar a}}   \left(\frac{\hbar a}{\sqrt{2} m c}\right)^{1/3} \sum_{s'=1}^2 \int_{\mathbb{R}^3} d^3 k \nonumber \\
& \times  \tilde{\psi}^*_{s'}(\vec{k}) \tilde{u}^\dagger_{s'}(\vec{k})  \tilde{\mathfrak{W}}_{\text{R} s}(\Omega, \vec{K}_\perp) \int_{\mathbb{R}^3} d^3 x \theta(\epsilon-|az-1| ) e^{-i\vec{k} \cdot \vec{x}} \nonumber \\
 &  \times \text{Ai} \left( \left(\frac{\sqrt{2} m c}{\hbar a}\right)^{2/3} \left(\left(\frac{\hbar |\vec{K}_\perp|}{\sqrt{2} m c}\right)^{2} + a z  - \frac{\hbar \Omega}{m c^2}  \right) \right).
\end{align}
The nonrelativistic condition (\ref{nonrelativistic_wavefunction}) in Eq.~(\ref{chi_6}) leads to
\begin{align}\label{chi_7}
& \chi_{\text{R} s}(\Omega,\vec{K}_\perp) \approx \frac{1-i}{(2 \pi)^{5/3}} \sqrt{\frac{m}{\hbar a}}   \left(\frac{\hbar a}{\sqrt{2} m c}\right)^{1/3}  \nonumber \\
& \times \sum_{s'=1}^2 \int_{\mathbb{R}^3} d^3 k \theta \left( \epsilon^{1/2}-\frac{\hbar |\vec{k}|}{mc} \right) \tilde{\psi}^*_{s'}(\vec{k}) \tilde{u}^\dagger_{s'}(\vec{k})  \tilde{\mathfrak{W}}_{\text{R} s}(\Omega, \vec{K}_\perp) \nonumber \\
& \times  \int_{\mathbb{R}^3} d^3 x \theta(\epsilon-|az-1| ) e^{-i\vec{k} \cdot \vec{x}} \nonumber \\
 &  \times \text{Ai} \left( \left(\frac{\sqrt{2} m c}{\hbar a}\right)^{2/3} \left(\left(\frac{\hbar |\vec{K}_\perp|}{\sqrt{2} m c}\right)^{2} + a z  - \frac{\hbar \Omega}{m c^2}  \right) \right).
\end{align}

One can now use the limits of the Airy function to show that if $\hbar \Omega/mc^2 - 1 \ll -\epsilon$, then $\chi_{\text{R} s}(\Omega,\vec{K}_\perp)$ is exponentially vanishing \cite{falcone2022}. Conversely, if $\hbar \Omega/mc^2 - 1 \gg \epsilon $, the Airy function appearing in Eq.~(\ref{chi_7}) oscillates way faster than $e^{-ik_3 z}$\cite{falcone2022} leading to a vanishing integration with respect to $z$. Consequently, $\chi_{\text{R} s}(\Omega,\vec{K}_\perp)$ is nonvanishing only when $|\hbar \Omega/mc^2 - 1 | \lesssim \epsilon$ (i.e., $\Omega$ is positive and nonrelativistic). As a result, $\hat{\mathfrak{C}}^\dagger_\psi$ only creates a nonrelativistic right-Rindler particle and can be approximated by the following identity
\begin{align}\label{C_Bogoliubov_transformations_4}
& \hat{\mathfrak{C}}_\psi \approx \sum_{s=1}^2  \int_\mathbb{R} d\Omega \theta \left( \epsilon - \left| \frac{\hbar \Omega}{mc^2} - 1 \right| \right)  \int_{\mathbb{R}^2} d^2 K_\perp\nonumber \\
& \times  \theta \left( \epsilon^{1/2}-\frac{\hbar |\vec{K}_\perp|}{mc} \right) \chi_{\text{R} s}(\Omega,\vec{K}_\perp) \hat{C}_{\text{R} s}(\Omega,\vec{K}_\perp).
\end{align}

Equation (\ref{C_Bogoliubov_transformations_4}) can be plugged in Eq.~(\ref{single_particle}) to give a Rindler-Fock representation of the Minkowski single particle. The operator $\hat{\mathfrak{C}}^\dagger_\psi$ creates a nonrelativistic right-Rindler single particle over the Minkowski vacuum background $|0_\text{M}\rangle$, which, in turns, is given by Eq.~(\ref{0_M_0_LR}). 

By using Eq.~(\ref{0_M_0_LR}) and the fact that $e^{-\pi \Omega / c a}$ is exponentially vanishing when $\Omega / c a \gg 1$, we find that the Minkowski vacuum is mainly populated by Rindler particles with frequency $\Omega \lesssim c a$. Such particles are localized far from the region $|aZ|\lesssim \epsilon$. Indeed, by using Eqs.~(\ref{UV_UV_tilde}), (\ref{W_tilde_W_tilde}) and (\ref{Bessel_limits_a}) one can prove that the modes representing right-Rindler particles and antiparticles with $\Omega \lesssim c a$ are vanishing in $|aZ|\lesssim \epsilon$. The result is that Eq.~(\ref{single_particle}) is approximately a nonrelativistic right-Rindler single particle created over a sea of Rindler particles and antiparticles localized far from $|aZ|\lesssim \epsilon$.

\section{Uniform electric field in the accelerated frame}

Here, we consider an electric field $\vec{E}$ that appears uniform in the inertial frame (i.e., $\vec{E}(t,\vec{x})=E \vec{e}_z$) and we compute the consequent electromagnetic field in the accelerated frame.

The electromagnetic tensor in the inertial frame is
\begin{equation}
F^{\mu \nu}=\frac{E}{c} \begin{pmatrix}
& 0 & 0 & 0 & 1 \\
& 0 & 0 & 0 & 0\\
& 0 & 0 & 0 & 0\\
& -1 & 0 & 0 & 0
\end{pmatrix}.
\end{equation}

The coordinate transformation $(t,\vec{x}) \mapsto (T,\vec{X})$ from the Minkowski to the right-Rindler frame is such that
\begin{subequations}
\begin{align}
& c a t = e^{ a Z} \sinh(c a T), && x=X, && y=Y, \\ & a z = e^{ a Z} \cosh(c a T),
\end{align}
\end{subequations}
which can be inverted to give
\begin{subequations}\label{coordinate_transformation_inverse}
\begin{align}
& c a T = \tanh^{-1}\left( \frac{ct}{z} \right), && X=x, && Y=y, \\ & aZ = \frac{1}{2} \log ((az)^2 - (cat)^2).
\end{align}
\end{subequations}
From Eq.~(\ref{coordinate_transformation_inverse}) one can compute the following Jacobian matrix
\begin{equation}\label{Jacobian}
\frac{\partial X^\mu}{\partial x^\nu} = \begin{pmatrix}
& e^{- a Z} \cosh(caT) & 0 & 0 & c^{-1} e^{- a Z} \sinh(caT)\\
& 0 & 1 & 0 & 0\\
& 0 & 0 & 1 & 0\\
& c e^{- a Z} \sinh(caT) & 0 & 0 & e^{- a Z} \cosh(caT)
\end{pmatrix}.
\end{equation}

The components of the electromagnetic tensor in the noninertial frame $(T,\vec{X})$ can be obtained by using Eq.~(\ref{Jacobian}) as follows
\begin{equation}\label{em_tensor_Rindler}
\frac{\partial X^\mu}{\partial x^\alpha} \frac{\partial X^\nu}{\partial x^\beta} F^{\alpha \beta} = e^{- 2 a Z} \frac{E}{c} \begin{pmatrix}
& 0 & 0 & 0 & 1 \\
& 0 & 0 & 0 & 0\\
& 0 & 0 & 0 & 0\\
& -1 & 0 & 0 & 0
\end{pmatrix}.
\end{equation}
Equation (\ref{em_tensor_Rindler}) states that in the accelerated frame, $\vec{E}$ appears as an electric field along $Z$ with magnitude $e^{- 2 a Z} E$. The consequent potential energy $V_\text{ext}$ that is vanishing for $Z = 0$ is reported by Eq.~(\ref{V_ext}).

\section{Spin degeneracy of accelerated hydrogen-like atom}

Here, we consider the energy potential $V_\text{ext}$ and the special and general relativity corrections to the accelerated hydrogen-like atom. We show that the spin degeneracy of the first energy level is not lifted.

The full Rindler-Dirac equation with potentials $V_\text{nuc}$ and $V_\text{ext}$ describing the electron in the accelerated frame is
\begin{equation} \label{Dirac_perturbed}
i \hbar \partial_0  \Psi  = H   \Psi,
\end{equation}
with the following Hamiltonian  (see, for instance, Ref.~\cite{falcone2022non})
\begin{align}\label{Dirac_Hamiltonian}
H = & - i \hbar c^2 \gamma^0 \gamma^3 \partial_3 - \frac{i}{2} \hbar \alpha \gamma^0 \gamma^3  + e^{aZ}\gamma^0(- i \hbar c^2  \gamma^1 \partial_1 \nonumber \\
& - i \hbar c^2 \gamma^2 \partial_2 + m c^3 ) + V_\text{nuc} + V_\text{ext}.
\end{align}

Notice that $H$ is symmetric with respect to the following unitary operators
\begin{align}
& U_1 = i c \gamma^0 P_1 \gamma^2 \gamma^3, & U_2 = i c \gamma^0 \gamma^1 P_2 \gamma^3,
\end{align}
where $P_i$ is the parity operator for the $i$-th coordinate, i.e., $P_1: X \mapsto -X$ and $P_2: Y \mapsto -Y$. Indeed, one can prove that $H$ commutes with $U_1$ and $U_2$,
\begin{align}\label{U_H_commutation}
& [U_1, H] = 0, & [U_2, H] = 0,
\end{align}
by using the following anticommutative properties
\begin{subequations}\label{P_gamma_identities}
\begin{align}
& \{ P_1, \partial_1 \} = 0, & \{ P_2, \partial_2 \} = 0,\label{P_identities} \\ 
& \{ \gamma^\mu, \gamma^\nu \} = - 2 \eta^{\mu\nu}.\label{gamma_identities}
\end{align}
\end{subequations}
Equation (\ref{gamma_identities}), the hermiticity of $\gamma^0$,  $P_1$ and $P_2$ and the antihermiticity of $\gamma^i$ can be used to prove the unitarity of $U_1$ and $U_2$. Furthermore, Eq.~(\ref{gamma_identities}) leads to the following anticommutative relation
\begin{equation}\label{U_anticommutation}
\{ U_1, U_2 \} = 0.
\end{equation}

Equation (\ref{U_H_commutation}) implies that $H$ and $U_1$ are simultaneously diagonalizable. The same occurs for $H$ and $U_2$. However, $U_1$ and $U_2$ are not compatible, since, as a consequence of Eq.~(\ref{U_anticommutation}), they do not commute.

Consider a state $\Psi$ which is simultaneously eigenstate of $H$ and $U_2$. The non-compatibility between $U_1$ and $U_2$ implies that $\Psi$ is not eigenstate of $U_1$. Hence, $U_1 \Psi$ is a different state from $\Psi$. However, $U_1 \Psi$ is still eigenstate of $H$ with the same energy of $\Psi$. We find that the Hamiltonian $H$ is at least two-degenerate for each energy level.

\bibliography{bibliography}